\documentclass[final]{IEEEtran}

\usepackage{cite}
\usepackage{threeparttable}
\usepackage{graphicx}
\usepackage{picinpar}
\usepackage[cmex10]{amsmath}
\usepackage{amsmath,amsfonts,amssymb}
\usepackage{subfigure}
\usepackage{algorithm}
\usepackage{algorithmic}
\usepackage{stfloats}
\usepackage{bm}
\usepackage{amsthm}
\usepackage{color}

\begin{document}
\newtheorem{lemma}{Lemma}
\newtheorem{corol}{Corollary}
\newtheorem{theorem}{Theorem}
\newtheorem{proposition}{Proposition}
\newtheorem{definition}{Definition}
\newcommand{\e}{\begin{equation}}
\newcommand{\ee}{\end{equation}}
\newcommand{\eqn}{\begin{eqnarray}}
\newcommand{\eeqn}{\end{eqnarray}}
\title{Asymptotic Orthogonality Analysis of Time-Domain Sparse Massive MIMO Channels}

\author{Zhen Gao, Linglong Dai, Chau Yuen, and Zhaocheng Wang
 \thanks{Z. Gao, L. Dai, and Z. Wang are with Tsinghua National Laboratory for
 Information Science and Technology (TNList), Department of Electronic Engineering,
 Tsinghua University, Beijing 100084, China (E-mails: gao-z11@mails.tsinghua.edu.cn; \{daill, zcwang\}@tsinghua.edu.cn).} %
\thanks{C. Yuen is with Singapore University of Technology and Design, Singapore
138682 (e-mail: yuenchau@sutd.edu.sg).} %
\thanks{This work was supported in part by the National Key Basic Research Program of China (Grant No. 2013CB329203), Singapore A*STAR SERC Project (Grant No. 142 02 00043), National Natural Science Foundation of China (Grant Nos. 61201185 and 61271266), and Beijing Natural Science Foundation (Grant No. 4142027).}  %

}

\maketitle
\begin{abstract}
The theoretical analysis of downlink massive MIMO usually assumes the ideal Gaussian channel matrix with asymptotic orthogonality of channel vectors associated with different users, since it can provide the favorable propagation condition.
Meanwhile, recent experiments have shown that massive MIMO channels between a certain user and massive base station antennas appear the spatial common sparsity (SCS) in both the time domain and angle domain. This motivates us to investigate whether realistic sparse massive MIMO channels could provide the favorable propagation condition, and reveal the capacity gap between massive MIMO systems over realistic sparse channels and that under the ideal Gaussian channel matrix assumption. 
This paper theoretically proves that channel vectors associated with different users in massive MIMO over realistic sparse channels satisfy the asymptotic orthogonality, which indicates that the favorable propagation condition can also be provided. Moreover, the simulation results confirm the theoretical analysis. 
\end{abstract}

\IEEEpeerreviewmaketitle
\section{Introduction}
The theoretical performance of massive MIMO relies on the favorable propagation condition, which is equivalent to the asymptotic orthogonality of channel vectors associated with different users~\cite{scaleMIMO}. For theoretical analysis, the channel is usually modeled as the ideal Gaussian channel matrix with elements following the mutually independent Gaussian distribution, so that channel vectors associated with different users are asymptotically orthogonal and can provide the favorable propagation condition for massive MIMO~\cite{scaleMIMO,WCL}.


Recent researches have shown that, due to the finite number of significant scatterers and the compact antenna array at the base station (BS), realistic massive MIMO channels between a certain user and massive base station (BS) antennas exhibit the intrinsically spatial common sparsity (SCS) \cite{{space_con},{outdoor},{my3},{SCS},{my2},{my1}}.
However, in \cite{{SCS},{my2},{my1}}, the proposed channel estimation schemes only consider the massive MIMO channels with SCS in the time domain, while \cite{{space_con}} and \cite{{my3}} only consider the SCS of massive MIMO channels in the angle domain.
 For massive MIMO channels, both the path delay in the time domain and angle of
departure (AoD)/angle of arrival (AoA) in the angle domain are essential parameters to represent
the channels.
Therefore, in this paper, we will jointly consider the SCS of massive MIMO channels in both the time domain and angle domain.
Clearly, the actual massive MIMO channel experienced by practical systems is different from the ideal Gaussian channel matrix for theoretical analysis. Such difference motivates us to investigate the propagation condition of realistic massive MIMO channels with SCS, and reveal the capacity gap between massive MIMO systems over actual sparse channels and that under the ideal Gaussian channel matrix assumption.
Specifically, we first introduce the SCS of massive MIMO channels in both the time domain and angle domain,
and illustrate the connection between the sparse massive MIMO channel model and the frequency-domain channel matrix model in massive MIMO-OFDM systems. Then, we prove that channel vectors over sparse channels are asymptotically orthogonal, where our analysis is based on the widely used isotropic uniform linear array (ULA) without loss of generality \cite{{space_con},{outdoor},{my3}}. The asymptotic orthogonality of different channel vectors indicates that massive MIMO channels with SCS could provide the favorable propagation condition, which is also confirmed by simulations. 

\section{System Model}
In typical massive MIMO systems, the BS employs $M$ antennas to simultaneously serve $K$ single-antenna users. Usually, $M$ is large, and much larger than $K$, e.g., $M=128$ and $K=16$~\cite{scaleMIMO}. In the downlink, the BS transmits data ${\bf{x}}\in {\mathbb{C}}^{M \times 1}$ to $K$ single-antenna users, and the received signal ${\bf{y}}\in {\mathbb{C}}^{K \times 1}$ for $K$ users can be expressed as
\begin{equation}\label{equ:downlink}
{\bf{y}} = {\sqrt \rho  _d}{\bf{Gx}} + {\bf{n}} = {\sqrt \rho  _d}{{\bf{D}}^{1/2}}{\bf{Hx}} + {\bf{n}},
\end{equation}
where $\rho_d$ is the transmit power, ${\bf{G}}\in {\mathbb{C}}^{K \times M}$ is the downlink channel matrix, ${\bf{n}} \in {\mathbb{C}}^{K \times 1}$ is the additive white Gaussian noise (AWGN) vector, the diagonal matrix ${\bf{D}}^{1/2}\in {\mathbb{C}}^{K \times K}$ with the $k$th diagonal element $\sqrt {{\beta _k}} $ represents the large-scale channel fading due to shadow fading and path loss, while ${\bf{H}} \in {\mathbb{C}}^{K \times M}$ denotes the small-scale channel fading matrix due to channel variation. Note that when the widely used OFDM is considered, (\ref{equ:downlink}) is valid for every frequency-domain subcarrier.

To fully exploit advantages of massive MIMO, the BS requires the downlink channel information for precoding, user scheduling, etc. The propagation condition of downlink channel matrix determines the achievable capacity of downlink massive MIMO, which can be expressed as~\cite{scaleMIMO}
 \begin{equation}\label{equ:C1}
{C_{{\rm{}}}} = {\log _2}\det \left( {{{\bf{I}}_K} + \rho_d {\bf{G}}{{\bf{G}}^*}} \right),
\end{equation}
where ${{\bf{I}}_K}\in {\mathbb{C}}^{K \times K}$ is the unit matrix, and $(\cdot )^{\rm{*}}$ is the Hermitian transpose operation. 
It can be observed from (\ref{equ:C1}) that the achievable capacity heavily depends on the property of channel matrix $\bf{G}$.

For the ideal Gaussian channel matrix, ${\bf{G}}$ provides the favorable propagation condition, where elements of the small-scale channel fading matrix $\bf{H}$ follow the independent and identically distributed (i.i.d.) circularly symmetric complex Gaussian distribution ${\cal{CN}}(0,1)$. It has been shown that row vectors of $\bf{G}$ are asymptotically orthogonal~\cite{scaleMIMO}, i.e.,
\begin{equation}\label{equ:asy}
\mathop {\lim }\limits_{M \to \infty } {\bf{G}}{{\bf{G}}^*}/M = \mathop {\lim }\limits_{M \to \infty } {{\bf{D}}^{1/2}}{\bf{H}}{{\bf{H}}^*}{{\bf{D}}^{1/2}}/M = {\bf{D}}.
\end{equation}
Under such ideal Gaussian channel matrix assumption, the available capacity has the simple asymptotic form as follows
\begin{equation}\label{equ:C2}
{\log _2}\det \left( {{{\bf{I}}_K} + {\rho _d}{\bf{G}}{{\bf{G}}^*}} \right)\mathop  \approx \limits^{M \gg K} \sum\limits_{k = 1}^K {{{\log }_2}\left( {1 + {\rho _d}M{\beta _k}} \right)} .
\end{equation}
Such favorable propagation condition can ensure that the inter-user interference vanishes when ${M \to \infty }$~\cite{scaleMIMO}. Accordingly, the superiority of massive MIMO in spectral efficiency can be guaranteed. For practical massive MIMO with a finite $M$, the favorable propagation condition indicates that the condition number of $\bf{G}$ should be as small as possible \cite{scaleMIMO}.

\section{Asymptotic Orthogonality Analysis for Massive MIMO Channels with SCS}
We first illustrate the realistic massive MIMO channels with SCS in both the time domain and angle domain. Then we theoretically prove that channel vectors are asymptotically orthogonal, where the first and second-order moments of the inner product of different channel vectors are derived.

\subsection{Massive MIMO Channels with SCS}
Due to the finite number of significant scatterers, for the time-domain channel impulse response (CIR), the majority of channel gains are concentrated on a small number of channel taps although the delay spread can be large in typical wireless communication systems \cite{{outdoor}}. The sparse downlink time-domain CIR between the $k$th user and the $m$th BS antenna can be expressed as \cite{SCS}
\begin{equation}\label{equ:CIR2}
{h_{km}}\left( t \right)\! = \!\sum\nolimits_{{s_{km}} = 1}^{{S_{km}}} \!{\alpha _{km}^{{s_{km}}}\delta \left( {t\! - \!\tau _{km}^{{s_{km}}}} \right)} ,~{S_{km}}\ll {{{\tau _{\max }}}}{{{f_s}}},
\end{equation}
where ${\alpha _{km}^{{s_{km}}}}$ and $ \tau _{km}^{{s_{km}}}$ are the ${s_{km}}$th path gain and path delay, respectively, ${S_{km}}$ is the number of resolvable multipaths, ${\tau _{\max }}$ is the maximum delay spread, $f_s$ is the system bandwidth, and ${{{\tau _{\max }}}}{{{f_s}}}$ is the normalized maximum delay~spread.

Moreover, there are measurements showing that time-domain CIRs between different co-located BS antennas and a certain user share very similar path delays \cite{scaleMIMO}, i.e., $ {s_{km}}={s_k} $ and $ {S_{km}}={S_k}$, 
for $0\le m\le M-1$. This is because the scale of the compact antenna array at the BS is small compared with the large distance of signal transmission, which leads to the fact that channels of different transmit-receive antenna pairs share very similar scatterers. 
The inherent sparsity of time-domain massive MIMO channels discussed above is referred to as the \emph{SCS in the time domain}.

Furthermore, for a certain multipath from one scatterer, the AoD seen from $M$ antennas at the BS are very similar, which is referred to as the \emph{SCS in the angle domain} of massive MIMO channels \cite{{outdoor},{space_con},{my3}}.
Due to such SCS in the angle domain, the path gains associated with $M$ antennas via the $s_k$th scatterer can be expressed as the steering vector, i.e.,
\begin{equation}\label{equ:steer}
\begin{small}
\begin{array}{l}
\left[ {\alpha _{k0}^{{s_{k0}}},\alpha _{k1}^{{s_{k1}}}, \cdots ,\alpha _{k\left( {M - 1} \right)}^{{s_{k\left( {M - 1} \right)}}}} \right] = \\
\alpha _k^{{s_k}}\left[ {1,{e^{j\frac{{2\pi d\sin (\theta _k^{{s_k}})}}{\lambda }}},{e^{j\frac{{2\pi 2d\sin (\theta _k^{{s_k}})}}{\lambda }}}, \cdots ,{e^{j\frac{{2\pi \left( {M - 1} \right)d\sin (\theta _k^{{s_k}})}}{\lambda }}}} \right],
\end{array}
\end{small}
\end{equation}
where the isotropic ULA with the antenna spacing $d$ is considered without loss of generality, $\alpha _{k0}^{{s_{k0}}}=\alpha^{{s_k}}_k$ is the path gain of the first ($m=0$) BS antenna, $ \theta^{{s_k}}_k$ is the AoD of the ${s_{k}}$th propagation path, and $\lambda$ denotes the wavelength.

The channel model with SCS in both the time domain and angle domain is different from the ideal Gaussian channel matrix with the inherent favorable propagation condition widely used for theoretical analysis. This motivates us to investigate whether the realistic sparse massive MIMO channels meet the favorable propagation condition, and reveal the capacity gap between massive MIMO over realistic sparse channels and that under the ideal Gaussian channel matrix assumption.

\subsection{Asymptotic Orthogonality Analysis}\label{asy}

According to (\ref{equ:CIR2}), the channel frequency response over the $n$th subcarrier, denoted by ${g_{km}}[n] $, can be expressed as
{\setlength\abovedisplayskip{1pt plus 1pt minus 3pt}
\setlength\belowdisplayskip{6pt plus 0pt minus 1pt}
\begin{equation}\label{equ:CIR3}
\begin{array}{l}
{g_{km}}[n] = \!\!\!\sum\limits_{{s_{km}} = 1}^{{S_{km}}} \!{\alpha _{km}^{{s_{km}}}\exp \left( { - \frac{{j2\pi n{f_s}\tau _{km}^{{s_{km}}}}}{N}} \right)} \\
 {\kern 28pt} =\!\!\! \sum\limits_{{s_{km}} = 1}^{{S_{km}}}\! {\alpha _{km}^{{s_{km}}}\exp \left( { - j2\pi {\gamma _n}\tau _{km}^{{s_{km}}}} \right)}, {\kern 3pt}1\le n\le N,
\end{array}
\end{equation}
}where $N$ is the size of the OFDM symbol and~${\gamma _n} = n{f_s}/N$.

Eq. (\ref{equ:CIR3}) builds the connection between time-domain massive MIMO channels and frequency-domain channel matrix.
For multi-user MIMO-OFDM systems, $g_{km}[n]$ is the element of the $k$th row and the $m$th column in the channel matrix ${\bf{G}}[n]$ over the $n$th subcarrier (see $\bf{G}$ in Eq. (\ref{equ:downlink})). Here ${\bf{G}}[n] = {\left[ ({{{\bf{g}}_1}{{[n]})^T},({{\bf{g}}_2}{{[n]})^T}, \cdots ,({{\bf{g}}_K}{{[n]})^T}} \right]^T}$ and ${{\bf{g}}_k}[n] = \left[ {{g_{k1}}[n],{g_{k2}}[n], \cdots ,{g_{kM}}[n]} \right]$.
Based on (\ref{equ:CIR3}), we will investigate the asymptotic orthogonality of row vectors of the frequency-domain channel matrix over sparse massive MIMO channels.




Due to the SCS in the angle domain as shown in (\ref{equ:steer}), for the ${s_{k}}$th propagation path, the path gain associated with the $m$th BS antenna $\alpha _{km}^{{s_{km}}}$ can be expressed as
\begin{equation}\label{equ:gain}
\alpha _{km}^{{s_{km}}} = \alpha _k^{{s_k}}{e^{j2\pi md\sin (\theta _k^{{s_k}})/\lambda }},{\kern 3pt} 0 \le m \le M-1.
\end{equation}
Furthermore, due to the SCS in the time domain, under the far field assumption, the corresponding path delay of the $m$th BS antenna via the $s_k$th scatterer can be expressed as \cite{SCS}
\begin{equation}\label{equ:delay}
\tau _{km}^{{s_{km}}} = \tau_k ^{{s_k}} + md\sin (\theta_k ^{{s_k}})/c, {\kern 3pt}0 \le m \le M-1,
\end{equation}
where $\tau _k^{{s_k}} $ is the path delay of the first ($m=0$) BS antenna, and $c$ is the velocity of electromagnetic waves.

Consequently, by substituting (\ref{equ:gain}) and (\ref{equ:delay}) into (\ref{equ:CIR3}), we have
\begin{equation}\label{equ:g}
\begin{small}
\begin{array}{l}
{g_{km}}[n] = \sum\limits_{{s_k} = 1}^{{S_k}} {\alpha _k^{{s_k}}{e^{j2\pi \frac{{md\sin (\theta_k ^{{s_k}})}}{\lambda }(1 - \frac{{\lambda {\gamma _n}}}{c}) - j2\pi {\gamma _n}\tau_k ^{{s_k}}}}} \\
  {\kern 31pt}= \sum\limits_{{s_k} = 1}^{{S_k}} {\alpha _k^{{s_k}}{e^{j2\pi \frac{{md\sin (\theta_k ^{{s_k}})}}{\lambda }(1 - \frac{{\lambda {\gamma _n}}}{c})}}\mu _{nk}^{{s_k}}},
\end{array}
\end{small}
\end{equation}
where $\mu _{nk}^{{s_k}} = {e^{ - j2\pi {\gamma _n}\tau_k ^{{s_k}}}}$.

%

For massive MIMO systems, since the distance between different users simultaneously communicating with the BS is usually large, their respective CIRs are usually mutually independent~\cite{scaleMIMO}. Meanwhile, we assume that their AoDs follow the i.i.d. uniform distribution ${\cal{U}}\left[ {0,2\pi } \right)$ \cite{3GPP}.
Due to the limited dominant scatterers, the limited AoDs appear the SCS.
Then, based on (\ref{equ:g}) and according to the definition of asymptotic orthogonality (\ref{equ:asy}), we have
\begin{equation}\label{equ:g2}
\begin{small}
\begin{array}{l}
\!\!\!\!\!\frac{{{{\bf{g}}_{{p}}}[n]{{\bf{g}}_{{q}}}{{[n]}^*}}}{M}= \frac{1}{M}\sum\limits_{m = 0}^{M-1} \left( {}\!\!\right.{\sum\limits_{{s_{_{{p}}}} = 1}^{{S_{{p}}}} {\alpha _{{p}}^{{s_{_{{p}}}}}\mu _{np}^{{s_{{p}}}}{e^{j2\pi \frac{{md\sin (\theta _{{p}}^{{s_{{p}}}})}}{\lambda }(1 - \frac{{\lambda {\gamma _n}}}{c})}}} } \\
~~~~~~~~~~~~~ \times \sum\limits_{{s_{_{{q}}}} = 1}^{{S_{{q}}}} {(\alpha _{{q}}^{{s_{_{{q}}}}}\mu _{nq}^{{s_{{q}}}})^*{e^{-j2\pi \frac{{md\sin (\theta _{q}^{{s_{{q}}}})}}{\lambda }(1 - \frac{{\lambda {\gamma _n}}}{c})}}}\left. {\!\!} \right),
\end{array}
\end{small}
\end{equation}
where $p$ and $q$ denote two arbitrary users with $1 \le p < q \le K$. Based on (\ref{equ:g2}), we will derive the first and second-order moments of ${{{\bf{g}}_p}[n]{{\bf{g}}_q}{{[n]}^*}/M}$.
Taking the statistical expectation of (\ref{equ:g2}) with respect to AoDs, we obtain
\begin{equation}\label{equ:g3}
\begin{small}
\begin{array}{l}
{E_\theta }\left\{ {{{\bf{g}}_p}[n]{{\bf{g}}_q}{{[n]}^*}/M} \right\}\\
= \frac{1}{M}\sum\limits_{m = 0}^{M - 1} {\sum\limits_{{s_p} = 1}^{{S_p}} {{\alpha _p^{{s_p}}}\mu _{np}^{{s_p}}\frac{1}{{2\pi }}\int\limits_0^{2\pi } {{e^{  j2\pi \frac{{md\sin (\theta _{p}^{{s_p}})}}{\lambda }(1 - \frac{{\lambda {\gamma _n}}}{c})}}d\theta _{p}^{{s_p}}} } } \\
~~~~~~~~~~~ \times \sum\limits_{{s_q=1}}^{{S_q}} {{{\left( {{\alpha _q^{{s_q}}}\mu _{nq}^{{s_q}}} \right)}^*}\frac{1}{{2\pi }}\int\limits_0^{2\pi } {{e^{-j2\pi \frac{{md\sin (\theta _{q}^{{s_q}})}}{\lambda }(1 - \frac{{\lambda {\gamma _n}}}{c})}}d\theta _{q}^{{s_q}}} } \\
 =  \sum\limits_{{s_p} = 1}^{{S_p}}{\sum\limits_{{s_q} = 1}^{{S_q}} {{z^{s_p}_{np}}(z^{s_q}_{nq})^*} } \frac{1}{M}\sum\limits_{m = 0}^{M - 1} {{J_0}{{\left( {2\pi \frac{{md}}{\lambda }(1 - \frac{{\lambda {\gamma _n}}}{c})} \right)}^2}}\\
 =\frac{{\upsilon _{pqn}}}{M}\sum\limits_{m = 0}^{M - 1} {{J_0}{{\left( {am} \right)}^2}} ,
\end{array}
\end{small}
\end{equation}
where ${E_{\bf{\theta }}}\left\{\cdot  \right\}$ represents the expectation operator for random variables (RVs) $\left\{ {\theta _{{p}}^{{s_{{p}}}}} \right\}_{{s_{{p}}} = 1}^{{S_{{p}}}}$ and $\left\{ {\theta _{{q}}^{{s_{{q}}}}} \right\}_{{s_{{q}}} = 1}^{{S_{{q}}}}$ subjected to the i.i.d. ${\cal{U}}\left[ {0,2\pi } \right)$, ${J_0}\left( {\cdot} \right)$ is the zero-order Bessel function of the first kind with the definition ${J_0}\left( x \right) = \frac{1}{{2\pi }}\int_0^{2\pi } {{e^{ \pm jx\sin \theta }}d\theta } $ \cite{Bessel}, ${z_{np}^{s_p}}{\rm{ = }}\alpha _p^{{s_p}}\mu _{np}^{{s_p}}$, ${z_{nq}^{s_q}}{\rm{ = }}\alpha _q^{{s_q}}\mu _{nq}^{{s_q}}$, $a = 2\pi \frac{d}{\lambda }(1 - \frac{{\lambda {\gamma _n}}}{c})$, and ${\upsilon _{pqn}} = {\sum\nolimits_{{s_p} = 1}^{{S_P}} {\sum\nolimits_{{s_q} = 1}^{{S_q}} {\alpha _p^{{s_p}}\mu _{np}^{{s_p}}{{\left( {\alpha _q^{{s_q}}\mu _{nq}^{{s_q}}} \right)}^*}} } }$. Besides, due to $f_c \gg f_s$ in typical massive MIMO systems, we have $\frac{{\lambda {\gamma _n}}}{c} = \frac{{n{f_s}}}{{N{f_c}}}\ll 1$ for $1 \le n \le N$ and $a>0$, where $f_c$ is the carrier frequency. According to (\ref{equ:g3}), the first-order moment of ${{{\bf{g}}_p}[n]{{\bf{g}}_q}{{[n]}^*}/M}$ can be expressed as
\begin{equation}\label{equ:bound2.1}
\!\!\!\!\begin{small}
\begin{array}{l}
E\!\left\{\!\! {\frac{{{{\bf{g}}_p}[n]{{\bf{g}}_q}{{[n]}^*}}}{M}}\!\! \right\}\! =\! {E_\alpha }\!\left\{\!\! {{E_\theta }\!\!\left\{ \!\! {\frac{{{{\bf{g}}_p}[n]{{\bf{g}}_q}{{[n]}^*}}}{M}} \!\!\right\}}\!\! \right\}\! =\! \frac{{E_\alpha }\left\{ \!{{\upsilon _{pqn}}}\! \right\}{\sum\limits_{m = 0}^{M - 1} \!\!{{J_0}{{\left( {am} \right)}^2}\!} }}{M},
 \end{array}
\end{small}
\end{equation}
where ${E}\left\{\cdot  \right\}$ is the expectation operator for all RVs, ${E_{\bf{\alpha }}}\left\{\cdot  \right\}$ is the expectation operator for RVs $\left\{ {\alpha _{{p}}^{{s_{{p}}}}} \right\}_{{s_{{p}}} = 1}^{{S_{{p}}}}$ and $\left\{ {\alpha _{{q}}^{{s_{{q}}}}} \right\}_{{s_{{q}}} = 1}^{{S_{{q}}}}$, and path gains and AoDs are mutually independent.

Since the large-scale channel fading has no impact on the asymptotic orthogonality of channel vectors, without loss of generality, we consider channels with normalized energy, i.e., $\sum\nolimits_{{s_p} = 1}^{{S_p}} {{{\left| {\alpha _{p}^{{s_p}}} \right|}^2}} \!  = \! 1$ and $\sum\nolimits_{{s_q} = 1}^{{S_q}} {{{\left| {\alpha _{q}^{{s_q}}} \right|}^2}}\!  = \! 1$. Then we have
\begin{equation}\label{equ:bound}
\begin{small}
\begin{array}{l}
{\left| {{\upsilon _{pqn}}} \right| \le \sum\limits_{{s_p} = 1}^{{S_p}} {\sum\limits_{{s_q} = 1}^{{S_q}} {\left| {\alpha _p^{{s_p}}\mu _{np}^{{s_p}}} \right|\left| {{{(\alpha _q^{{s_q}}\mu _{nq}^{{s_q}})}^*}} \right|}  \le \sqrt {{S_p}{S_q}} } },
 \end{array}
\end{small}
\end{equation}
where the second inequality utilizes the Cauchy-Schwarz inequality \cite{Bessel}.


Consequently, based on the discussion above, as well as the boundedness ($\left| {{J_0}\left( x \right)} \right| < 1$ for $x>0$) and the delay of ${J_0}\left( x \right)$ proportional to $1/\sqrt{x}$ when $x>0$ increases \cite{Bessel}, we have
\begin{equation}\label{equ:bound2}
\begin{small}
\begin{array}{l}
\!\!\!\!\mathop {\lim }\limits_{M \to \infty } \left| {E\left\{ {\frac{{{{\bf{g}}_p}[n]{{\bf{g}}_q}{{[n]}^*}}}{M}} \right\}} \right| = \mathop {\lim }\limits_{M \to \infty } \frac{\left| {{E_\alpha }\left\{ {{\upsilon _{pqn}}} \right\}} \right|{\sum\limits_{m = 0}^{M - 1} {{J_0}{{\left( {am} \right)}^2}} }}{M}\\\!\!\!\!
\le \!\!\mathop {\lim }\limits_{M \to \infty }\!\!\!\!\! \frac{{{E_\alpha }\left\{  \left|{{\upsilon _{pqn}}} \right|\right\}}{\sum\limits_{m = 0}^{M - 1} {{J_0}{{\left( {am} \right)}^2}} }}{M}
\!\!\!\le \!\!\!\sqrt {{S_p}{S_q}} \!\mathop {\lim }\limits_{M \to \infty }\!\!\!\! \frac{{\sum\limits_{m = 0}^{M - 1}\!\! {{J_0}{{\left( {am} \right)}^2}} }}{M} \!\!= \!0,
\end{array}
\end{small}
\end{equation}
where the first inequality is based on Jensen's inequality.

Moreover, we investigate the second-order moment of ${{{\bf{g}}_p}[n]{{\bf{g}}_q}{{[n]}^*}/M}$, namely,
\begin{equation}\label{equ:bound2.3}
\begin{small}
\begin{array}{l}
\!\!\!\!\!\!{\rm{var}}\left\{ {\frac{{{{\bf{g}}_p}[n]{{\bf{g}}_q}{{[n]}^*}}}{M}} \right\}{\rm{ = }}E\left\{ {\frac{{{{\left| {{{\bf{g}}_p}[n]{{\bf{g}}_q}{{[n]}^*}} \right|}^2}}}{{{M^2}}}} \right\}{\rm{ - }}E{\left\{ {\frac{{{{\bf{g}}_p}[n]{{\bf{g}}_q}{{[n]}^*}}}{M}} \right\}^2},
 \end{array}
\end{small}
\end{equation}
where ${\rm{var}}\left\{\cdot  \right\}$ is the variance of a RV. Furthermore, we have
\begin{equation}\label{equ:bound3}
\begin{small}
\begin{array}{*{20}{l}}
{\frac{{{{\left| {{{\bf{g}}_p}[n]{{\bf{g}}_q}{{[n]}^*}} \right|}^2}}}{{{M^2}}} = \frac{{\sum\limits_{m = 0}^{M - 1} {\sum\limits_{{s_p} = 1}^{{S_p}} {\sum\limits_{{s_q} = 1}^{{S_q}} {z_{np}^{{s_p}}{e^{jam\sin (\theta _p^{{s_p}})}}{{\left( {z_{nq}^{{s_q}}{e^{jam\sin (\theta _q^{{s_q}})}}} \right)}^*}} } } }}{M}}\\
~~~~~~~~{ \times \frac{{\sum\limits_{m' = 0}^{M - 1} {\sum\limits_{{{s}_p'} = 1}^{{S_p}} {\sum\limits_{{{s}_q'} = 1}^{{S_q}} {z_{nq}^{{{s}_q'}}{e^{jam'\sin (\theta _q^{{{s}_q'}})}}{{\left( {z_{np}^{{{s}_p'}}{e^{jam'\sin (\theta _p^{{{s}_p'}})}}} \right)}^*}} } } }}{M},}
\end{array}
\end{small}
\end{equation}
and we can obtain its expectation for AoDs as follows
\begin{equation}\label{equ:bound4}
\begin{small}
\!\!\!\begin{array}{*{20}{l}}
{{E_\theta }\!\!\left\{ \!\!{\frac{{{{\left| {{{\bf{g}}_p}[n]{{\bf{g}}_q}{{[n]}^*}}\right|}^2}}}{{{M^2}}}\!\! } \right\}\!\! = \!\! \frac{1}{{{M^2}}}\!\!\!\!\!\sum\limits_{m = 0}^{M - 1} \!{\sum\limits_{m' = 0}^{M - 1} \! \! \!{\left\{ \!{\sum\limits_{{s_p} = 1}^{{S_p}} \!{\sum\limits_{{s_q} = 1}^{{S_q}}  \!\!{{{\left| {z_{np}^{{s_p}}} \right|}^2}\!{{\left| {z_{nq}^{{s_q}}} \right|}^2}\!\!{J_0}\!\!\left( {a(m{\rm{ - }}m')} \right)^2} } } \right.} } }\\
{{\rm{ + }}\!\!\!\sum\limits_{{s_p} = 1}^{{S_p}} \! {\sum\limits_{{s_q} = 1}^{{S_q}} {\sum\limits_{{s_q'} = 1,{s_q'} \ne {{s}_q}}^{{S_q}} \!\!\!\!\!\!\!{{{\left| {z_{np}^{{s_p}}} \right|}^2}\!{{\left( {z_{nq}^{{s_q}}} \right)}^{\rm{*}}}z_{nq}^{{{s}_q'}}{J_0}\! \left( {a(m{\rm{ - }}m')} \right)\!{J_0}\! \left( {am} \right)\!{J_0}\! \left( {am'} \right)} } } }\\
{ + \!\!\!\sum\limits_{{s_q} = 1}^{{S_q}} \!{\sum\limits_{{s_p} = 1}^{{S_p}}{\sum\limits_{{s_p'} = 1,{s_p'} \ne {{s}_p}}^{{S_p}} \!\!\!\!\!\!\!\! {{{\left| {z_{nq}^{{s_q}}} \right|}^2}\!{{\left( {z_{np}^{{{s}_p'}}} \right)}^{\rm{*}}}\!\!z_{np}^{{s_p}}{J_0}\! \left( {a(m'{\rm{ - }}m)} \right)\!{J_0}\! \left( {am} \right)\!{J_0}\! \left( {am'} \right)} } } }\\
{ + \!\!\!\!\!\sum\limits_{{s_p} = 1}^{{S_p}} \!{\sum\limits_{{s_p'} = 1,{s_p'} \ne {{s}_p}}^{{S_p}}\! {\sum\limits_{{s_q} = 1}^{{S_q}} \!{\sum\limits_{{s_q'} = 1,{s_q'} \ne {{s}_q}}^{{S_q}} \!\!\!\!\!{{{\left(\!\! {z_{np}^{{{s}_p'}}} \!\!\right)}^{\rm{*}}}} } } } \!\!\!z_{np}^{{s_p}}\!\!\left. {{{\left(\! {z_{nq}^{{s_q}}}\! \right)}^{\rm{*}}}\!\!z_{nq}^{{{s}_q'}}\!{J_0}{{\left( {am} \right)}^2}{J_0}{{\left( {am'} \right)}^2}} \!\!\right\}.}
\end{array}
\end{small}
\end{equation}
Similar to (\ref{equ:bound}), we can get the following inequality
\begin{equation}\label{equ:bound5}
\begin{small}
\begin{array}{l}
\!\!\!\!\!\!\!E\left\{\!\! {\frac{{{{\left| {{{\bf{g}}_p}[n]{{\bf{g}}_q}{{[n]}^*}}\! \right|}^2}}}{{{M^2}}}}\!\! \right\}\!\!= \!\!{E_\alpha }\!\!\!\left\{ \!{{E_\theta  }\!\!\left\{ \!\!{\frac{{{{\left| {{{\bf{g}}_p}[n]{{\bf{g}}_q}{{[n]}^*}} \!\right|}^2}}}{{{M^2}}}}\!\! \right\}} \!\!\right\} \!\!\le \!\! \frac{1}{{{M^2}}}\!\!\!\!\sum\limits_{m = 0}^{M - 1} \!\!{\sum\limits_{m' = 0}^{M - 1}\! \!\! {\left\{ \!\!{{J_0}\! \left(\!{a(m{\rm{\!  - \! }}m')} \! \right)^2} \right.} } \\
~~~{\rm{ + }}({S_q}{\rm{ + }}{S_p}{\rm{ - 2}}){J_0}\left( {a(m{\rm{ - }}m')} \right){J_0}\left( {am} \right){J_0}\left( {am'} \right)\\
~~~\left. { + ({S_p}{\rm{ - }}1)({S_q} - 1){J_0}{{\left( {am} \right)}^2}{J_0}{{\left( {am'} \right)}^2}} \right\}.
\end{array}
\end{small}
\end{equation}
Similar to (\ref{equ:bound2.1}) and (\ref{equ:bound2}), we can further obtain
\begin{equation}\label{equ:bound5.1}
\begin{small}
\begin{array}{l}
\!\!\!\!\!\!\mathop {\lim }\limits_{M \to \infty } E\left\{ {\frac{{{{\left| {{{\bf{g}}_p}[n]{{\bf{g}}_q}{{[n]}^*}} \right|}^2}}}{{{M^2}}}} \right\}{\rm{ = }}\mathop {\lim }\limits_{M \to \infty } {E_\alpha }\left\{ {{E_\theta }\left\{ {\frac{{{{\left| {{{\bf{g}}_p}[n]{{\bf{g}}_q}{{[n]}^*}} \right|}^2}}}{{{M^2}}}} \right\}} \right\}{\rm{ = }}0.
\end{array}
\end{small}
\end{equation}
Then we have $\mathop {{\rm{lim}}}\limits_{M \to \infty } {\rm{var}}\left\{ {{{\bf{g}}_p}[n]{{\bf{g}}_q}{{[n]}^*}/M} \right\} \!\!= 0$ by substituting (\ref{equ:bound2}) and (\ref{equ:bound5.1}) into (\ref{equ:bound2.3}). Consequently, the asymptotic orthogonality
of row vectors of the channel matrix ${\bf{G}}[n]$ is proven.

It should be pointed out that we can further exploit the statistical characteristics of path gains to simplify (\ref{equ:bound2.1}) and (\ref{equ:bound2.3}). Specifically, if we consider the $S_k$-path channels with $\left\{ {\alpha _{{k}}^{{s_{{k}}}}} \right\}_{{s_{{k}}} = 1}^{{S_{{k}}}}$ following the i.i.d. ${\cal{CN}}(0,1/{S_k})$, we have $E\left\{ {\sum\nolimits_{{s_k} = 1}^{{S_k}} {{{\left| {\alpha _{}^{{s_k}}} \right|}^2}} } \right\} = 1$, ${E_\alpha }\left\{ {{\upsilon _{pqn}}} \right\}=0$, $E\left\{ {z_{np}^{{s_p}}} \right\} = 0$, and $E\left\{ {z_{nq}^{{s_q}}} \right\} = 0$. Therefore, (\ref{equ:bound2.1}) and (\ref{equ:bound2.3}) can be further expressed as
\begin{equation}\label{equ:bound6}
\begin{small}
\begin{array}{l}
E\left\{ {{{\bf{g}}_p}[n]{{\bf{g}}_q}{{[n]}^*}/M} \right\}{\rm{ = }}\sum\limits_{m = 0}^{M - 1} {{J_0}{{\left( {am} \right)}^2}{E_\alpha }\left\{ {{\upsilon _{pqn}}} \right\}} /M = 0,
\end{array}
\end{small}
\end{equation}
\begin{equation}\label{equ:bound7}
\begin{small}
\begin{array}{l}
{\rm{var}}\left\{ {{{\bf{g}}_p}[n]{{\bf{g}}_q}{{[n]}^*}/M} \right\}{\rm{ = }}{E_\alpha }\left\{ {{E_\theta }\left\{ {{{\left| {{{\bf{g}}_p}[n]{{\bf{g}}_q}{{[n]}^*}} \right|}^2}/{M^2}} \right\}} \right\}\\
~~~~~~~~~~~~~~~~~~~~~~~~~~~{\rm{ = }}\sum\limits_{m = 0}^{M - 1} {\sum\limits_{m' = 0}^{M - 1} {{J_0}\left( {a(m{\rm{ - }}m')} \right)^2} }/M^2.
\end{array}
\end{small}
\end{equation}

\section{Numerical Results}\label{section_sim}

In simulations, the widely adopted ULA at the BS was considered, system carrier was $f_c=2~\rm{GHz}$, system bandwidth was $f_s=10~\rm{MHz}$, the OFDM size was $N= 4096$, and the guard interval length was $N_g=\tau_{\rm{max}}f_s=64$ to combat the maximum channel delay spread of $\tau_{\rm{max}} = 6.4~\mu s$ \cite{3GPP}. Besides, we adopted Rayleigh fading sparse multipath channels with typical $S=6$ paths model~\cite{3GPP}, where the normalized maximum delay spread is $N_g=64$ and path gains follows the i.i.d. ${\cal{CN}}(0,1/S)$. While for the ideal Gaussian channel matrix, the large-scale channel fading is removed (${\bf{D}}={\bf{I}}_K$), and elements of small-scale channel fading matrix $\bf{H}$ follow the i.i.d. ${\cal{CN}}(0,1)$.

\begin{figure}[!tp]
     \centering
     \includegraphics[width=9.1cm, keepaspectratio]
     {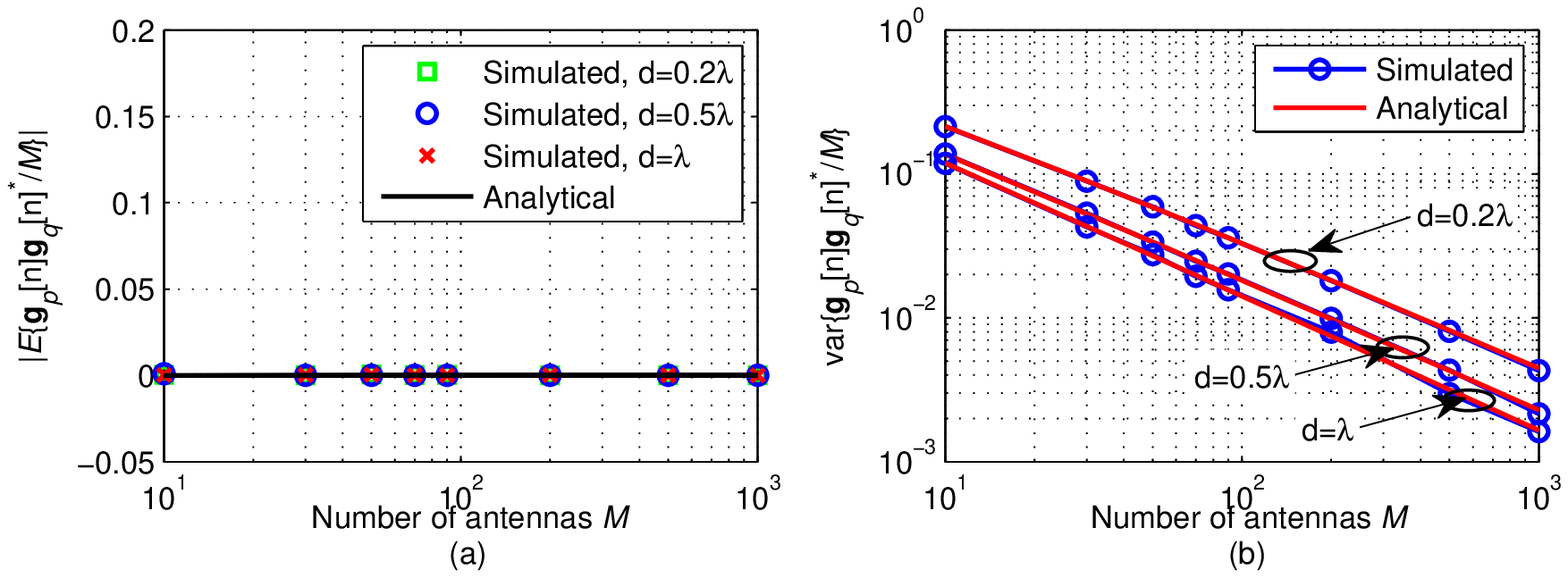}
    \caption{$\!|E\left\{ {{{\bf{g}}_p}[n]{{\bf{g}}_q}{{[n]}^*}/M} \right\}|$ and ${\rm{var}}\left\{ {{{\bf{g}}_p}[n]{{\bf{g}}_q}{{[n]}^*}/M} \right\}$ against $M$ and~$d$.}
     \label{fig:sim_ana}
\end{figure}

Fig. \ref{fig:sim_ana} shows the simulated and analytical $\!|E\left\{ {{{\bf{g}}_p}[n]{{\bf{g}}_q}{{[n]}^*}/M} \right\}|$ and ${\rm{var}}\left\{ {{{\bf{g}}_p}[n]{{\bf{g}}_q}{{[n]}^*}/M} \right\}$ against $M$ and $d$. Clearly, the analytical (\ref{equ:bound6}) and (\ref{equ:bound7}) have the excellent tightness with the simulation results, which confirms the validity of our theoretical analysis on the first and second-order moments of the inner product of channel vectors.
Moreover, ${\rm{var}}\{ {{{\bf{g}}_p}[n]{{\bf{g}}_q}{{[n]}^*}/M} \}$ decreases when $d$ increases, which implies that the larger $d$ can lead to the smaller inter-user interference.
Besides, both analytical and simulated curves confirm the asymptotic orthogonality of different channel vectors over sparse multipath channels.
%

\begin{figure}[!tp]
     \centering
     \includegraphics[width=7.6cm, keepaspectratio]
     {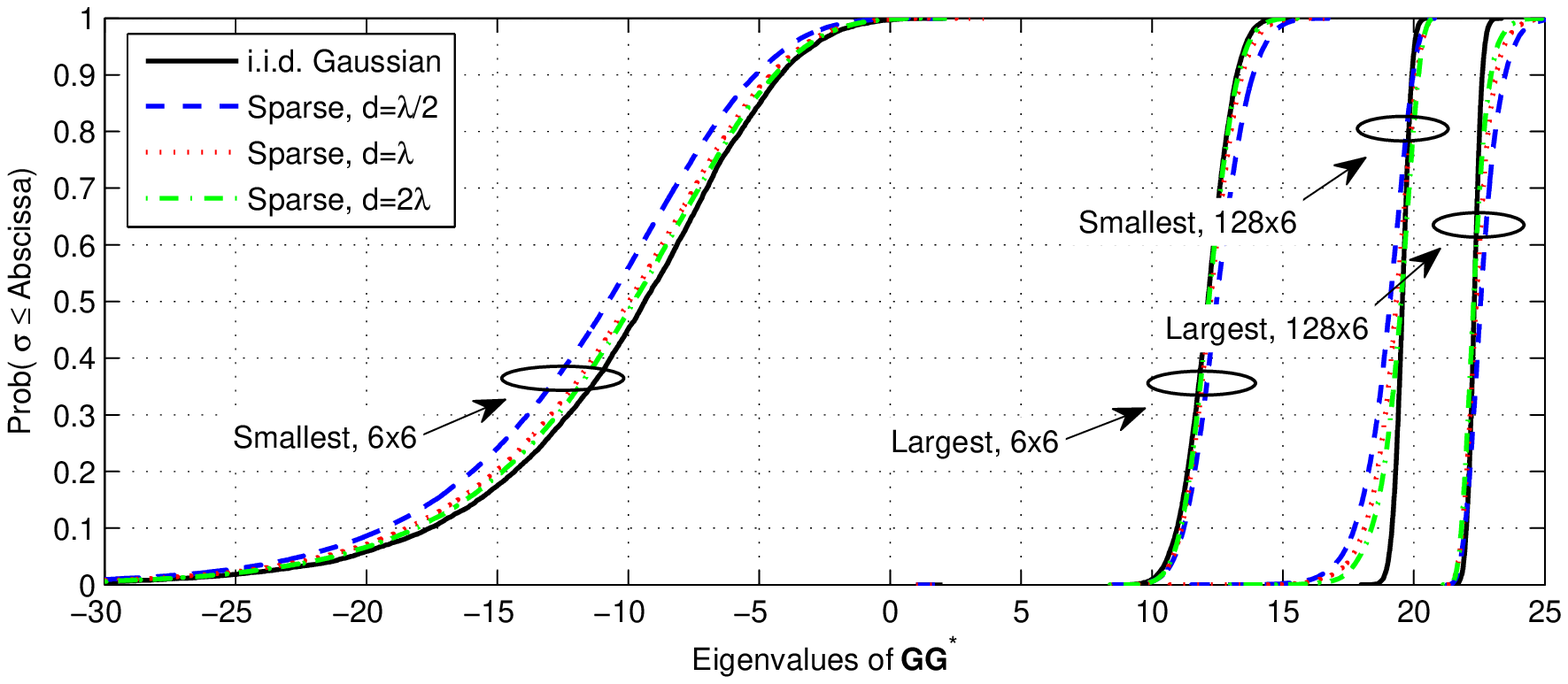}
    \caption{CDFs of smallest/largest eigenvalue of $\bf{G}\bf{G}^*$ in MIMO systems.}
     \label{fig:EV}
\end{figure}

Fig. \ref{fig:EV} provides the cumulative distribution function (CDF) of the smallest/largest eigenvalue of $\bf{G}\bf{G}^*$ for small-scale MIMO systems ($M=K=6$) and massive MIMO systems ($M=128$, $K=6$). Meanwhile, we plot the CDF of the smallest/largest eigenvalue of $\bf{G}\bf{G}^*$ under the ideal Gaussian channel matrix assumption as the benchmark for comparison.
Clearly, the gap of eigenvalue distributions of $\bf{G}\bf{G}^*$ over sparse multipath channels and that under the ideal Gaussian channel matrix is negligible, and this gap is narrowed when $d$ increases. Besides, the simulated CDFs of eigenvalue distributions of $\bf{G}\bf{G}^*$ over sparse multipath channels conform to the experimental results provided by Fig. 6 of \cite{scaleMIMO}. 

\begin{figure}[!tp]
     \centering
     \includegraphics[width=7.0cm, keepaspectratio]
     {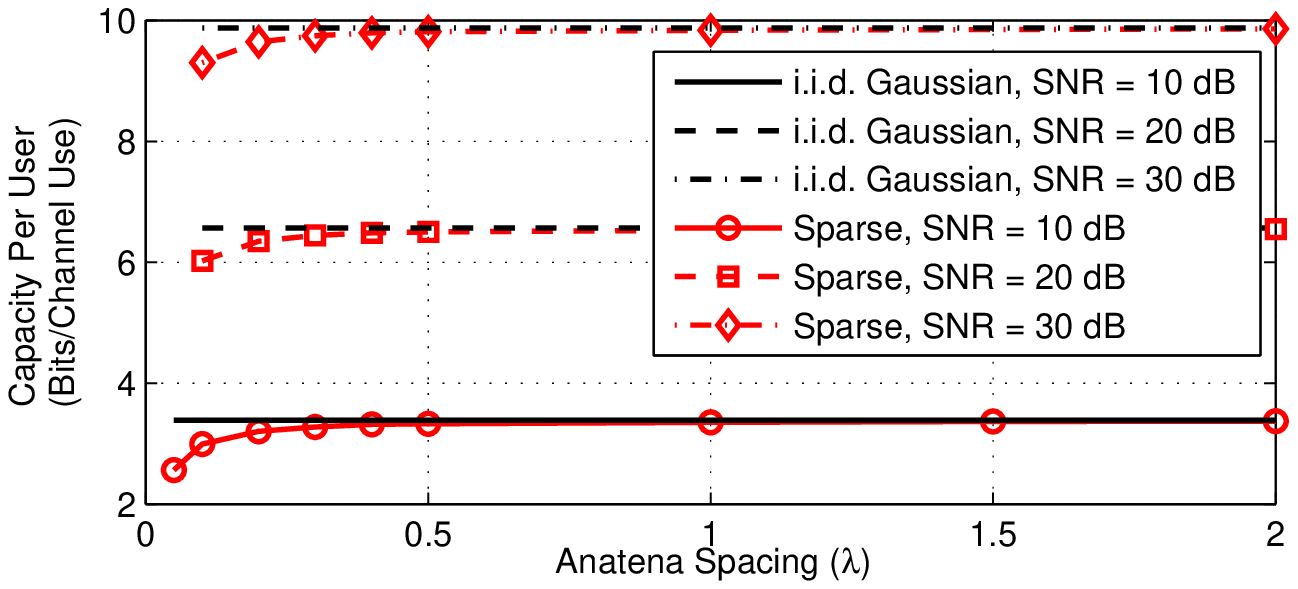}
    \caption{Comparison of capacity per user (bits/channel use) against the antenna spacing $\lambda$ in the massive MIMO system with $M=128$ and $K=16$.}
     \label{fig:Capacity}
\end{figure}

Fig. \ref{fig:Capacity} shows the capacity of massive MIMO systems with $\!\!\!M\!\!\!\!=\!\!\!128$ and $\!K\!\!\!=\!\!\!16$ over sparse multipath channels and that under the ideal Gaussian channel matrix assumption. 
The capacity gap between massive MIMO systems over realistic sparse multipath channels and that under the ideal Gaussian channel matrix assumption is small, and this gap is narrowed when $d$ becomes large, since the inter-user interference is reduced when $d$ becomes large as confirmed in Fig. \ref{fig:sim_ana}. Note that we do not consider the antenna coupling, which may further degrade the attainable capacity when $d$ is small. The small capacity loss conforms to the experimental results that the capacity of implemented massive MIMO always approaches the capacity under favorable propagation condition~\cite{scaleMIMO}. Besides, the small capacity loss implies that we may realize BS antenna array with small form factor (e.g., $\!d<\! 0.5 \lambda $) at the cost of a small capacity loss if the antenna coupling can be settled well.

%

\section{Conclusions}

This paper has investigated the propagation condition and system capacity of massive MIMO over realistic sparse channels with SCS in both time domain and angle domain. We have proven the equivalence between the favorable propagation condition desired by massive MIMO and realistic channels with SCS.
Specifically, we have derived the first and second-order moments of the inner product of different channel vectors over sparse multipath channels. Moreover, the asymptotic orthogonality of channel vectors over massive channels with SCS is proven. This proof indicates that the capacity of massive MIMO over realistic massive MIMO channels with SCS can approach that of the ideal Gaussian channel matrix assumption when the number of BS antennas is unlimited. The asymptotic orthogonality of channel vectors has been verified by comparing the simulated and analytical first and second-order moments of the inner product of channel vectors. Moreover, the simulated condition number of channel matrix and achievable capacity of massive MIMO systems over sparse MIMO channels are consistent with experimental results. The negligible capacity gap between massive MIMO over realistic sparse channels and that under the ideal Gaussian channel matrix assumption 
may inspire us to realize massive MIMO with better performance by exploiting the channel sparsity.

\begin{thebibliography}{10}




\bibitem{scaleMIMO}F. Rusek, {\it et al.}~
``Scaling up MIMO: Opportunities and challenges with very large arrays,"
{\em IEEE Signal Process. Mag.}, vol. 30, no. 1, pp. 40-60, Jan. 2013.

\bibitem{WCL}A. Pitarokoilis, S. K. Mohammed, and E. G. Larsson,
``On the optimality of single-carrier transmission in large-scale antenna systems,"
{\em IEEE Wireless Commun. Lett.}, vol. 1, no. 4, pp. 276-279, Aug. 2012.

















\bibitem{outdoor}T. Santos, J. Kredal, P. Almers, F. Tufvesson, and A. Molisch, ``Modeling
the ultra-wideband outdoor channel: Measurements and parameter extraction method,"
 {\it IEEE Trans. Wireless. Commun.}, vol. 9, no. 1, pp. 282-290, Jan. 2010.

\bibitem{SCS}Y. Barbotin and M. Vetterli, ``Estimation of sparse MIMO channels with
common support,"
 {\it IEEE Trans. Commun.}, vol. 60, no. 12, pp. 3705-3716, Dec. 2012.






\bibitem{my1}Z. Gao, L. Dai, C. Yuen, and Z. Wang,
``Super-resolution sparse MIMO-OFDM channel estimation based on spatial and temporal correlations,"
{\em IEEE Commun. Lett.}, vol. 18, no. 7, pp. 1266-1269, July 2014.

\bibitem{my2}C. Qi and L. Wu, ``Uplink channel estimation for massive MIMO systems exploring joint
channel sparsity,"
{\em Electron. Lett.}, vol. 50, no. 23, pp. 1770-1772, Nov. 2014.


\bibitem{space_con}C. Masouros and M. Matthaiou,
``Space-constrained massive MIMO: Hitting the wall of favorable propagation,"
 {\it IEEE Commun. Lett.}, vol. 19, no. 5, pp. 771-774, May 2015.

\bibitem{my3}N. Kolomvakis, M. Matthaiou, and M. Coldrey, ``Massive MIMO in sparse channels," in {\it Proc. IEEE SPAWC 2014} (Toronto, Canada), Jun. 2014. pp. 21-25.










\bibitem{3GPP}3GPP Technical Specification 36.211. Evolved Universal Terrestrial Radio Access (E-UTRA); Physical Channels and Modulation. 






\bibitem{Bessel}M. Abramowitz and I. A. Stegun,
{\it Handbook of Mathematical Functions with Formulas, Graphs, and Mathematical Tables}, U.S. Government Printing Office, 1964.



















\end{thebibliography}
\end{document}